\documentclass[twocolumn,aps,prb,superscriptaddress,floatfix]{revtex4-1}
\usepackage{graphicx}
\usepackage{dcolumn}
\usepackage[T1]{fontenc}
\usepackage{amsmath}
\usepackage{sidecap}

\begin{document}
\title{Determination of spin Hall angle in heavy metal/CoFeB-based heterostructures with interfacial spin-orbit fields}

\author{Witold Skowro\'{n}ski}
 \email{skowron@agh.edu.pl}
 \thanks{These authors contributed equally to this work}
\affiliation{AGH University of Science and Technology, Department of Electronics, Al. Mickiewicza 30, 30-059 Krak\'{o}w, Poland}
\author{\L{}ukasz Karwacki}
 \email{karwacki@ifmpan.poznan.pl}
  \thanks{These authors contributed equally to this work}
\affiliation{AGH University of Science and Technology, Department of Electronics, Al. Mickiewicza 30, 30-059 Krak\'{o}w, Poland}
\affiliation{Institute of Molecular Physics, Polish Academy of Sciences,
ul. Smoluchowskiego 17, 60-179 Pozna\'{n}, Poland}
\author{S\l{}awomir Zi\k{e}tek}
\affiliation{AGH University of Science and Technology, Department of Electronics, Al. Mickiewicza 30, 30-059 Krak\'{o}w, Poland}
\author{Jaros\l{}aw Kanak}
\affiliation{AGH University of Science and Technology, Department of Electronics, Al. Mickiewicza 30, 30-059 Krak\'{o}w, Poland}
\author{Stanis\l{}aw \L{}azarski}
\affiliation{AGH University of Science and Technology, Department of Electronics, Al. Mickiewicza 30, 30-059 Krak\'{o}w, Poland}
\author{Krzysztof Grochot}
\author{Tomasz Stobiecki}
\affiliation{AGH University of Science and Technology, Department of Electronics, Al. Mickiewicza 30, 30-059 Krak\'{o}w, Poland}
\affiliation{Faculty of Physics and Applied Computer Science, AGH University of Science and Technology,
30-059 Krak\'{o}w, Poland}
\author{Piotr Ku\'{s}wik}
\affiliation{Institute of Molecular Physics, Polish Academy of Sciences,
ul. Smoluchowskiego 17, 60-179 Pozna\'{n}, Poland}
\affiliation{Center for Advanced Technology, Adam Mickiewicz University, 89C
Umultowska Str., 61-614 Pozna\'{n}, Poland}
\author{Feliks Stobiecki}
\affiliation{Institute of Molecular Physics, Polish Academy of Sciences,
ul. Smoluchowskiego 17, 60-179 Pozna\'{n}, Poland}
\author{J\'{o}zef Barna\'{s}}
\affiliation{Institute of Molecular Physics, Polish Academy of Sciences,
ul. Smoluchowskiego 17, 60-179 Pozna\'{n}, Poland}
\affiliation{Faculty of Physics, Adam Mickiewicz University, ul. Umultowska 85, 61-614 Pozna\'{n},
Poland}

\begin{abstract}
Magnetization dynamics in W/CoFeB, CoFeB/Pt and W/CoFeB/Pt multilayers was investigated using spin-orbit-torque ferromagnetic resonance (SOT-FMR) technique. An analytical model based on magnetization dynamics  due to SOT was used to fit heavy metal (HM) thickness dependence of symmetric and antisymmetric components of the SOT-FMR signal. The analysis resulted in a determination of the properties of HM layers, such as spin Hall angle and spin diffusion length. The spin Hall angle of -0.36 and 0.09 has been found in the W/CoFeB and CoFeB/Pt bilayers, respectively, which add up in the case of W/CoFeB/Pt trilayer. More importantly, we have determined effective interfacial spin-orbit fields at both W/CoFeB and CoFeB/Pt interfaces, which are shown to cancel Oersted field for particular thicknesses of the heavy metal layers, leading to pure spin-current-induced dynamics and indicating the possibility for a more efficient magnetization switching.
\end{abstract}

\maketitle

\section{Introduction}
\label{sec:intro}
In heavy metal (HM) layers exhibiting significant spin-orbit coupling, the charge current ($j_\mathrm{c}$) may be converted into the spin current ($j_\mathrm{s}$) due to the spin Hall effect (SHE). The generated spin current, in turn, may exert a torque on the magnetization in an adjacent ferromagnet (FM) \cite{hirsch_spin_1999, sinova_universal_2004}. This phenomenon can be used, for instance, to control the magnetization state of next generation MRAM cells \cite{liu_spin-torque_2012, miron_perpendicular_2011} or to drive the magnetization precession in spin torque oscillators \cite{liu_magnetic_2012, demidov_magnetic_2012}. 
It has been already established, that in HM/FM bilayers the magnetization dynamics is driven by two components (damping-like and field-like) of the spin-orbit-torque (SOT)~\cite{kim_anomalous_2014} and by the Oersted field produced by the charge current. This effect is often used to quantitatively analyze the spin Hall angle $\theta$ = $j_\mathrm{s}$/$j_\mathrm{c}$~\cite{liu_spin-torque_2011}.  However, one can also expect the interface charge-spin conversion originating from the Rashba-type spin-orbit interactions \cite{miron_perpendicular_2011, kim_layer_2013, cecot_influence_2017} to play a significant role in such systems.
A strong interface effect has already been found in Ta/CoFeB bilayers~\cite{allen_experimental_2015,cecot_influence_2017} and recently in Ta/CoFeB/Pt trilayers~\cite{huang_engineering_2018} by analyzing the HM and FM thickness dependence of the SOT-FMR signal lineshape.

The above mentioned Rashba phenomenon at the interface (known also as the Edelstein-Rashba effect) has been modeled, among others, for a magnetized two-dimensional electron gas (2DEG) in both ballistic and diffusive regimes~\cite{Manchon2009,Ado,Dyrdal2017,Kim2017,Borge,Xiao}.  Theoretical results show that the SOT due to interfacial non-equilibrium (current-induced) spin polarization has symmetry similar to that induced by the spin Hall effect in heavy metals, i.e., there can be both damping-like and field-like components. However, in contrast to the spin Hall-induced torque and earlier mechanisms of spin-transfer torque (STT), the interfacial spin-orbit coupling (ISOC) acts as an effective field on the magnetization and, therefore, is not associated with transfer of the transverse part of spin current.  Moreover, the field-like component is mostly dominating, which can be attributed to a weak short-range spin-independent disorder~\cite{Borge}.

Another related interfacial effect that occurs at ferromagnet/heavy metal interfaces is the so-called spin Hall magnetoresistance effect (SMR)~\cite{althammer_quantitative_2013, avci_magnetoresistance_2015, Chen2016, Kim2016, Choi, manchon_new_2015}. 
In the bilayer under discussion, some of the electrons flowing from the HM into ferromagnet can have spin component parallel to the magnetization of the ferromagnetic layer -- due to external magnetic field, magnetic proximity effect, or magnetic anisotropy in the ferromagnet. This component of spin current is reflected from the interface and \textit{via} the inverse spin Hall effect (ISHE) in HM ~\cite{miao_inverse_2013} is converted to the charge current flowing parallel to the initial  current, which results in a reduced  resistance.
In contrast, the perpendicular component of spin current, that gives rise to spin torque exerted on the magnetization, is almost completely absorbed, and thus does not lead to charge current induced by ISHE. Although this effect occurs mostly near to the interface, it  is strongly dependent on the thickness of HM layer. It has been previously assumed that the origin of FMR signal in the case of HM/FM systems is the anisotropic magnetoresistance (AMR) of FM~\cite{liu_spin-torque_2012,allen_experimental_2015,taniguchi_spin-transfer_2015}, which might be the case for some systems. However, one should also take into account a contribution to the signal from SMR, as it has the same angular symmetry as AMR~\cite{Harder2016,Kim2016, Choi, cho_large_2015}.

\begin{figure}
\centering
\includegraphics[width=\columnwidth]{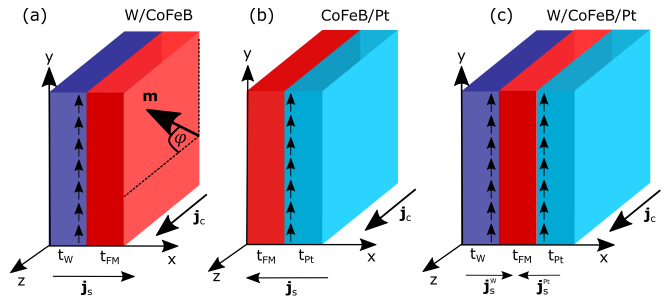}
\caption{Schematic representation of the structures examined in the paper: (a) W/CoFeB, (b) CoFeB/Pt, and (c) W/CoFeB/Pt. Spin currents, $\mathbf{j}_s$, and charge currents, $\mathbf{j}_c$, are indicated for all studied configurations. Angle 
$\varphi$ is the angle between magnetization $\mathbf{m}$ and direction of charge current, $\mathbf{j}_c$. Moreover, direction of spin accumulation at the interfaces is indicated.}
\label{fig:fig1}
\end{figure}

In this work, the SOT-FMR technique is used to investigate W/CoFeB and CoFeB/Pt bilayers, shown schematically in Fig.~\ref{fig:fig1}(a) and (b), as a function of thickness of HM ($t_\mathrm{HM}$). We have chosen W \cite{pai_spin_2012, skowronski_temperature_2017} and Pt \cite{hahn_comparative_2013, skinner_spin-orbit_2014, yang_layer_2015, sagasta_tuning_2016} as a source of spin current, since they are characterized by  the spin Hall angles of  opposite signs \cite{yu_large_2016, woo_enhanced_2014, bekele_dominancy_2018}. The evolution of symmetric and antisymmetric parts of the resonance signal with $t_\mathrm{HM}$ is fitted using the developed analytical model. As a result, the magnitudes of the effective magnetic fields associated with  damping-like and field-like components of SOT, as well as interface effects, such as SMR and ISOC, are determined. The model also enables evaluation of the spin Hall angle and spin diffusion length for each bilayer system. Finally, in the case of W/CoFeB/Pt trilayer, shown schematically in Fig.~\ref{fig:fig1}(c), we have used the values obtained from the constituent bilayers in order to identify the contributions from both heavy metals and their interfaces to the SOT induced in the trilayer system.

The paper is organized as follows: Section~\ref{sec:experiment} includes description of the experiment. Theoretical model, in turn, is presented in Sec.~\ref{sec:signal}, where the formulas for mixing voltage and spin Hall angle are derived and discussed. Experimental results are presented in Sec.~\ref{sec:results}, together with theoretical predictions based on the previous section. Finally, summary and concluding remarks are presented in Sec.~\ref{sec:summary}.  

\section{Experiment}
\label{sec:experiment}
Magnetron sputtering technique was used to deposit the following multilayer structures on Si/SiO$_\mathrm{2}$ substrates: W($t_\mathrm{W}$)/CoFeB(5)/Ta(1), CoFeB(5)/Pt($t_\mathrm{Pt}$) and W(5)/CoFeB(5)/Pt($t_\mathrm{Pt}$) (thicknesses in nm). The CoFeB layers were deposited from an alloy target with the composition of 20 at \% Co, 60 at \% Fe, and 20 at \% B. In case of W sputtering, a low DC power of 4 W and 6 cm target-sample distance was used, which resulted in deposition rate of 0.01 nm/s. Such conditions are essential for growth of thick W layers in highly resistive $\beta$-phase~\cite{hao_beta_2015,hao_giant_2015, neumann_temperature_2016}. The remaining materials were deposited with a 15 W DC power. For multilayers with a top material susceptible to oxidation, 1-nm thick Ta layer was deposited, which oxidized completely and formed a non-conducting protection layer. The thickness of wedges ranged from 0 to 10 nm in case of $t_\mathrm{Pt}$ and $t_\mathrm{W}$.

The bi- and trilayers were subsequently patterned into 100-$\mu$m long ($l$) and 20-$\mu$m wide ($w$) strips using electron-beam lithography and lift-off process with Al(10)/Au(50) contact pads. The resistivity of HM and FM was determined using the method described in Ref.~[\onlinecite{kawaguchi_anomalous_2018}] and the resistivity of FM, whose thickness was constant, was on average $\rho_{\text{FM}}\approx 104~\mu\Omega\text{cm}$.
The angular dependence of the resistance, enabling AMR and SMR determination, was measured at fixed magnetic field $H=500~\textrm{Oe}$ (which is sufficient to saturate the magnetization) applied at varying angle $\varphi$ with respect to the microstrip axis, using a custom-build rotating probe station. During the SOT-FMR measurements, an amplitude modulated radio-frequency (RF) current of the corresponding power of $P$ = 16 dBm and the frequency changing between 4 and 10 GHz was injected into the microstripe. The mixing voltage ($V_\mathrm{mix}$) was measured using lock-in amplifier synchronized to the RF signal. An in-plane magnetic field ($H$) applied at $\varphi$ = 30$^{\circ}$ with respect to the microstrip axis was swept from 0 up to 1250 Oe.

\section{Origin of the signal}
\label{sec:signal}

Mixing voltage generated in SOT-FMR experiment can be written down as time-averaged product of RF current with amplitude $I_{\text{RF}}$, $I(t)=\text{Re}\lbrace I_{\text{RF}}e^{i\omega t}\rbrace$, and time-dependent resistance, $R(t)$, of the system,
\begin{equation}
\label{eq:vmix0}
V_{\text{mix}}\equiv\langle V_{\text{mix}} (t)\rangle_t=\langle I(t)R(t) \rangle_t\,.
\end{equation}
We assume that resistance changes due to a combination of the AMR and SMR effects~\cite{Harder2016},
\begin{equation}
R(t)=R_0+\Delta R\cos^2{\varphi(t)},
\end{equation}
where $\varphi(t)$ is the time-dependent tilt-angle of the magnetization from its equilibrium orientation, $\Delta R\equiv \Delta R(\Delta R_{\text{AMR}},\Delta R_{\text{SMR}})$, and $R_0$ is the time-independent component of the resistance, which contains terms from both AMR and SMR. Here $\Delta R_{\text{AMR}}$ is the increment of anisotropic magnetoresistance of CoFeB assumed to be weakly dependent on the heavy metal thickness. In turn, the SMR contribution~\cite{Chen2016,Kim2017},
\begin{equation}
\label{eq:smr}
\Delta R_{\text{SMR}}\approx R_0^{\text{HM}}\theta_{\text{HM}}^2\frac{\lambda_{\text{HM}}}{t_{\text{HM}}}\tanh\frac{t_{\text{HM}}}{\lambda_{\text{HM}}}
\frac{g_{\text{HM}}^{R}}{1+g_{\text{HM}}^{R}}\tanh^2\frac{t_{\text{HM}}}{2\lambda_{\text{HM}}},
\end{equation}
is strongly dependent on the thickness $t_{\text{HM}}$ of HM layer. Furthermore,  $\theta_{\text{HM}}$ in Eq.~(\ref{eq:smr}) is the spin Hall angle of the HM, $\lambda_{\text{HM}}$ is the spin diffusion length in this material, and $g_{\text{HM}}^R=2e^2/\hbar\lambda_{\text{HM}}\rho_{\text{HM}}G_{\text{HM}}^R\coth{(t_{\text{HM}}/\lambda_{\text{HM}})}$ is the dimensionless real part of the spin-mixing conductance. In Eq.~(\ref{eq:smr}) we omitted the imaginary part of the spin-mixing conductance, as vanishingly small. The real part of spin-mixing conductance can be deduced from experiment according to the formula~\cite{Tserkovnyak}
\begin{align}
G_{\text{HM}}^{R}&=\frac{4\pi M_{s}t_{\textrm{FM}}}{\gamma \hbar}|\Delta \alpha|
 \,,
\end{align}
where $\Delta \alpha$ is the difference between Gilbert damping  coefficients $\alpha$ of pure CoFeB and CoFeB with W or Pt layers attached, $M_{s}$ denotes saturation magnetization of the HM/FM bilayer, and $\gamma$ is the gyromagnetic ratio. The relevant parameters have been collected in Table~\ref{tab:table1}. As $G_{\text{HM}}^{R}$ is derived experimentally, we treat it as an effective spin-mixing conductance, which also takes into account possible effects due to spin memory losses \cite{Qui_enhanced_2016}.

The mixing voltage can  be written down as follows:
\begin{equation}
\label{eq:vmix}
V_{\text{mix}}=-\frac{1}{4M_s}I_{\text{RF}}\Delta R\sin{(2\varphi_0)} \text{Re}\lbrace m_y\rbrace\,,
\end{equation}
where $\varphi_0$ is the equilibrium angle of magnetization (determined by applied magnetic field $H$) with respect to the direction of current, and $\text{Re}\lbrace m_y\rbrace$ is the $y$ component of magnetization vector found by solving Landau-Lifshitz-Gilbert (LLG) equation in the macrospin approximation,
\begin{equation}
\label{eq:llg}
\frac{\partial\mathbf{m}}{\partial t}-\alpha\mathbf{m}\times\frac{\partial \mathbf{m}}{\partial t}=\boldsymbol{\Gamma}\,.
\end{equation}
Here, $\mathbf{m}$ is a unit vector along the magnetization,  and
\begin{equation}
\boldsymbol{\Gamma}=-\gamma\mu_0\mathbf{m}\times\mathbf{H}_{\text{eff}}-\gamma\mu_0\mathbf{m}\times\mathbf{H}_{\text{ind}}
\end{equation}
determines the torques exerted on the magnetization due to effective magnetic field $\mathbf{H}_{\text{eff}}$ consisting of the demagnetization field, anisotropy field, and external magnetic field, and due to the current-induced field $\mathbf{H}_{\text{ind}}$.

The current-induced field $\mathbf{H}_{\text{ind}}$ consists of in-plane, $\mathbf{H}_{\parallel}$, and out-of-plane, $\mathbf{H}_{\perp}$, terms. In systems consisting of FM and HM layers, the only contribution to $\mathbf{H}_{\perp}$ comes from damping-like field, $\mathbf{H}_{\text{DL}}$, due to spin currents induced by the spin Hall effect in heavy metal layers.
The in-plane field, on the other hand, contains components due to Oersted field, $\mathbf{H}_{\text{Oe}}$, and interfacial spin-orbit field, $\mathbf{H}_{\text{so}}$.

The damping-like contributions to the effective field from W and Pt have the same sign due to opposite signs of the corresponding spin Hall angles,
\begin{equation}
\mathbf{H}_{\text{DL}}^{\text{HM}}=\pm H_{\text{DL}}^{\text{HM}}\mathbf{m}\times(\hat{\mathbf{x}}\times\hat{\mathbf{j}}_{c}^{\text{HM}}),
\end{equation}
where $+(-)$ corresponds to the spin current flowing from W (Pt) layer. The amplitude of damping-like field can be written in the following form:
\begin{equation}
\label{eq:hdl}
H_{\text{DL}}^{\text{HM}}=-\frac{\hbar j_{c}^{\text{HM}}}{2e\mu_0M_s^{\text{HM}}t_{\textrm{FM}}}\xi_{\text{DL}}^{\text{HM}}\,,
\end{equation}
where
\begin{align}
\xi_{\text{DL}}^{\text{HM}}\approx&\theta_{\text{HM}}\left(1-\text{sech}\frac{t_{\text{HM}}}{\lambda_{\text{HM}}}\right)
\frac{g_{\text{HM}}^{R}}{1+g_{\text{HM}}^{R}}
\end{align}
is the so-called damping-like spin Hall efficiency.

We also introduce the Oersted field,
\begin{equation}
\label{eq:hoe}
\mathbf{H}_{\text{Oe}}^{\text{HM}}=\pm\frac{1}{2}j_{c}^{\text{HM}}t_{\text{HM}}\hat{\mathbf{x}}\times\hat{\mathbf{j}}_{c}^{\text{HM}}\,,
\end{equation}
and the spin-orbit field,
\begin{equation}
\label{eq:hso}
\mathbf{H}_{\text{so}}^{\text{HM}}=\Gamma_{\text{so}}^{\text{HM}}\hat{\mathbf{x}}\times\hat{\mathbf{j}}_{c}^{\text{HM}}\,,
\end{equation}
with $\Gamma_{\text{so}}^{\text{HM}}$ being the amplitude of the effective spin-orbit field. In the following considerations we include the effective field corresponding to the  field-like torque induced by the spin Hall effect  into the spin-orbit field. Thus, the $\Gamma_{\text{so}}^{\text{HM}}$ amplitude contains both the interfacial Rashba-Edelstein and spin Hall field-like contributions.

\begin{figure*}
\centering
\includegraphics[width=\textwidth]{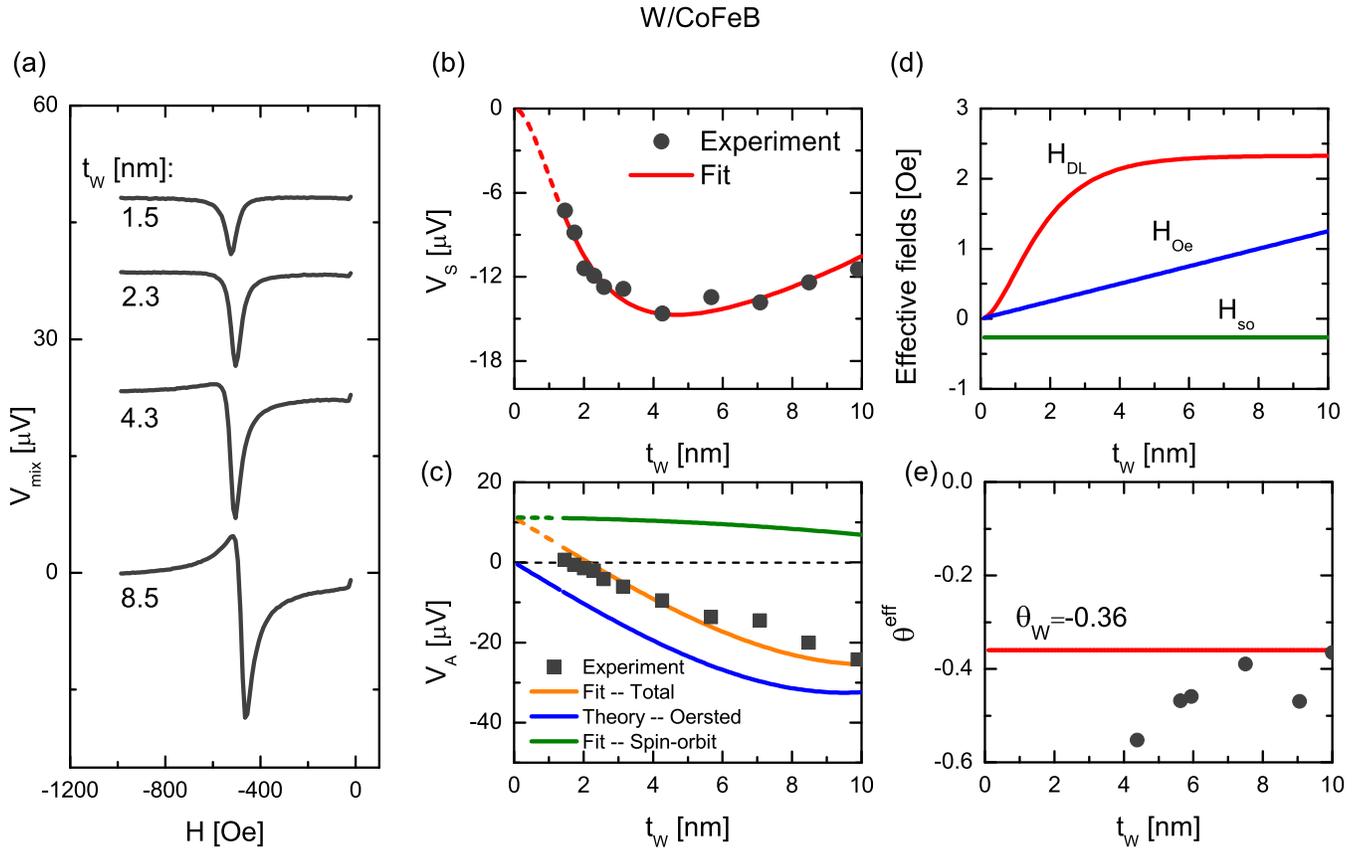}
\caption{DC mixing voltage, $V_{\text{mix}}$, measured in W/CoFeB microstripe as a function of magnetic field, $H$,  is shown in (a). The $V_{\text{mix}}$ curves for different $t_\mathrm{W}$ are artificially offset for clarity. Respective amplitudes of the symmetric and antisymmetric parts vs. $t_{\mathrm{W}}$, together with the fitted curves, are presented in (b) and (c). Solid lines show fitting result to the theoretical model. Calculated components of the effective field as a function of $t_\mathrm{W}$ are shown in (d). Experimental and theoretical values of the effective spin Hall angle, $\theta^{\text{eff}}$, are presented in (e). The dashed lines (red, green, and orange) represent interpolation of the model.}
\label{fig:w}
\end{figure*}

By linearizing the LLG equation~(\ref{eq:llg}) and inserting the obtained expression for $m_y$  into Eq.~(\ref{eq:vmix}), one obtains the following formula for the mixing voltage,
\begin{align}
V_{\text{mix}}=V_S\frac{\Delta H^2}{(H-H_0)^2+\Delta H^2}+V_A\frac{(H-H_0)\Delta H}{(H-H_0)^2+\Delta H^2}\,,
\end{align}
where $V_{\text{S}}$ is the amplitude of the symmetric part of the signal,
\begin{align}
V_S&=-\frac{1}{4}\frac{I_{\text{RF}}\Delta R\sin{(2\varphi_0)}2\pi f}{(2H+M_{\text{eff}})\gamma\mu_0}\frac{H_{\perp}}{\Delta H}
\end{align}
and $V_{\text{A}}$ is the amplitude of the antisymmetric component of mixing voltage,
\begin{equation}
V_A=-\frac{1}{4}\frac{I_{\text{RF}}\Delta R\sin{(2\varphi_0)}2\pi f}{(2H+M_{\text{eff}})\gamma\mu_0}\sqrt{1+\frac{M_{\text{eff}}}{H}}\frac{H_{\parallel}}{\Delta H}\,.
\end{equation}

To obtain the effective spin Hall angle of the structure one could use the ratio of symmetric and antisymmetric contributions to the mixing voltage, which yields
\begin{align}
&\frac{2e\mu_0M_st_{\textrm{FM}}}{\hbar}\frac{V_S}{V_A}\sqrt{1+\frac{M_{\text{eff}}}{H}} \nonumber \\
&=\frac{2e\mu_0M_st_{\textrm{FM}}}{\hbar}\frac{H_{\perp}}{H_{\parallel}}
\,.
\end{align}
This formula, however, does not give  the proper effective spin Hall angle (defined as $\theta^{\text{eff}} = j_s/j_c$), as it takes into account all the out-of-plane contributions. This formula can be rewritten as
\begin{align}
\frac{2e\mu_0M_st_{\textrm{FM}}}{\hbar}\frac{H_{\text{DL}}}{H_{\text{Oe}}+H_{\text{so}}}\,,
\end{align}
or equivalently
\begin{align}
\frac{2e\mu_0M_st_{\textrm{FM}}}{\hbar}\frac{H_{\text{DL}}}{H_{\text{Oe}}}\frac{1}{1+\frac{H_{\text{SO}}}{H_{\text{Oe}}}}\,.
\end{align}
Only assuming  that $H_{\text{so}}\ll H_{\text{Oe}}$, which is fulfilled for thick HM layers, one obtains the proper effective spin Hall angle, consistent with previous works (e.g. Ref.~[\onlinecite{liu_spin-torque_2012}]),
\begin{align}
\label{eq:eff_sha}
\frac{2e\mu_0M_st_{\textrm{FM}}}{\hbar}\frac{t_{\text{HM}}}{2}\frac{H_{\text{DL}}}{H_{\text{Oe}}}=\frac{j_s}{j_c}\equiv\theta^{\text{eff}}\,.
\end{align}

\section{Results and discussion}
\label{sec:results}

\begin{table}
\caption{\label{tab:table1}%
Parameters for W($t_{\text{W}}$)/CoFeB(5) and CoFeB(5)/Pt($t_{\text{Pt}}$) (layer thicknesses in nm). The last 3 rows include parameters obtained from fitting  the developed model to the experimental data.
}
\begin{ruledtabular}
\begin{tabular}{ l c c c}
 &
 W/CoFeB &
 CoFeB/Pt &
Units
 \\
 \colrule
$\rho_{\text{HM}}$ & 116 & 112 &$\mu\Omega\text{cm}$\\
$\alpha\times 10^{3}$ & 3.4 & 8.0 &  \\
$\alpha_{\textrm{CoFeB(5)}}\times 10^{3}$ & 4.0 & 4.0 \\
$|\Delta \alpha|\times 10^{3}$ & 0.6 & 4.0 &  \\
$\mu_0M_st_{\textrm{FM}}$ & 8  & 7.5& $\text{T}\text{nm}$ \\
$G^R\times 10^{-19}(e^2/\hbar)$ & 0.6 & 3.9& $\Omega^{-1}\text{cm}^{-2}$ \\
$j_{c}\times 10^{-10}$  &  2 & 2.9& $\text{A}/\text{m}^2$ \\
\colrule
$\theta$ & -0.36 & 0.09& \\
$\lambda$ & 1.3 & 2.2& nm \\
$\Gamma_{\text{so}}$ & -0.27 & 0.54& $\text{Oe}$
\\
\colrule
\end{tabular}
\end{ruledtabular}
\end{table}
The resistivity of each material was determined from the measured sheet conductance $G=l/(wR)$ of each microstripe as a function of $t_\mathrm{HM}$, according to the procedure described in Ref.~ [\onlinecite{kawaguchi_anomalous_2018}]. The resistivity of W was constant for the thicknesses above 2 nm: $\rho_\mathrm{W}$ = 116~$\mu \Omega$cm, which indicates the existence of the highly-resistive $\beta$-phase. In CoFeB/Pt bilayer the resistivity $\rho_\mathrm{Pt}$ = 112~$\mu \Omega$cm was determined. 
However, we found that the resisistivity of magnetron sputtering deposited Pt depends on whether Pt is deposited on crystalline underlayer (Co) or amorphous CoFeB alloy. In the first case, depending on the thickness of the bottom layer, the resistance was from about 20 to 100 $\mu\Omega\textrm{cm}$~\cite{sagasta_tuning_2016,kawaguchi_anomalous_2018,schoen_ultra-low_2016}, while in the second case from 100 to 200 $\mu\Omega\textrm{cm}$~\cite{Conca2017}.
\begin{figure*}
\centering
\includegraphics[width=\textwidth]{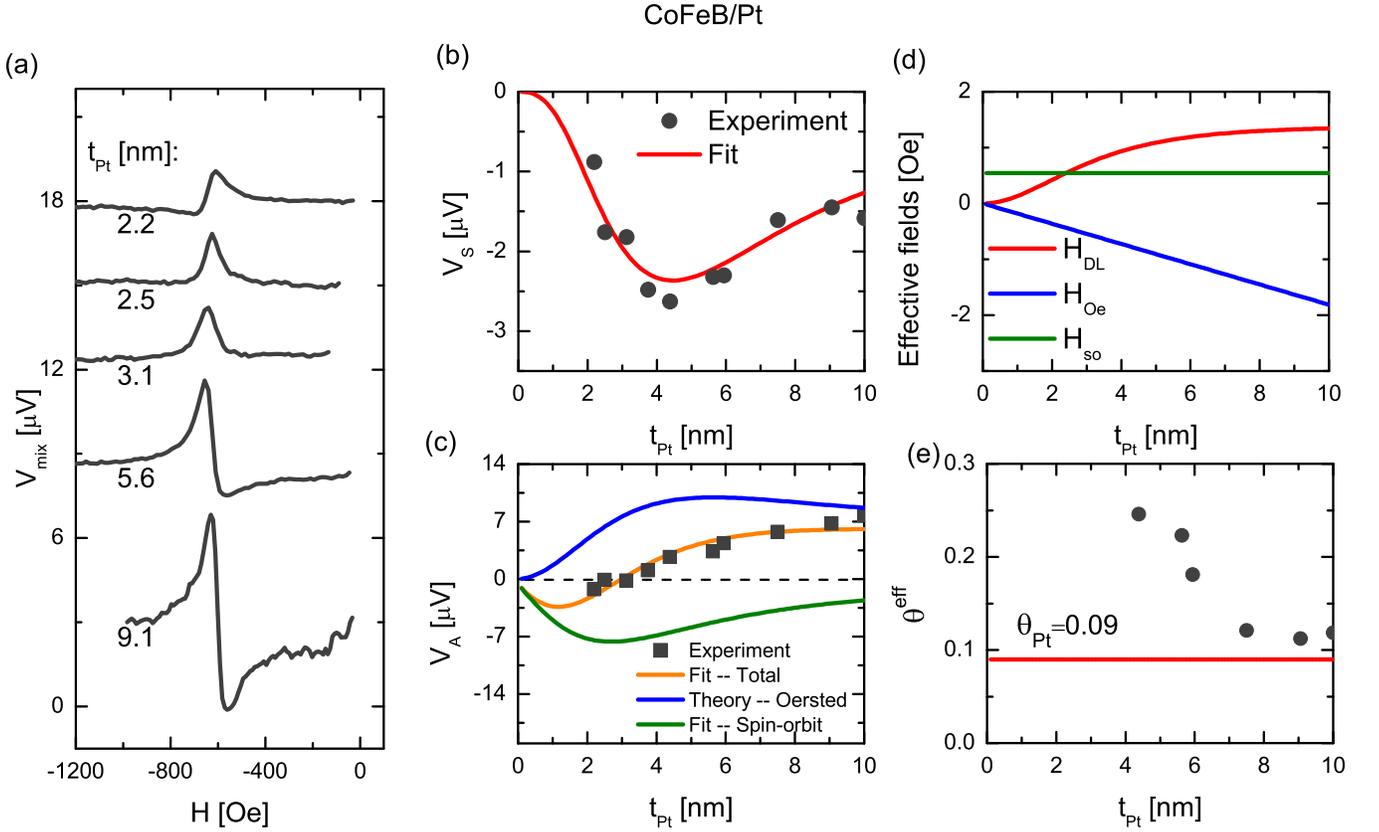}
\caption{DC mixing voltage, $V_{\text{mix}}$, measured in CoFeB/Pt microstripe as a function of magnetic field, $H$,  is shown in (a). The $V_{\text{mix}}$ curves for different $t_\mathrm{Pt}$ are artificially offset for clarity. Note that symmetry of the signal changes for $t_\mathrm{Pt}$ = 3 nm. Respective amplitudes of the symmetric and antisymmetric parts vs. $t_\mathrm{Pt}$, together with the fitted curves, are presented in (b) and (c). Calculated components of the effective field as a function of $t_\mathrm{Pt}$ are shown in (d). Experimental and theoretical values of the effective spin Hall angle, $\theta^{\text{eff}}$, are presented in (e).}
\label{fig:pt}
\end{figure*}
Now, we focus on the magnetization dynamics investigated by the SOT-FMR technique. Mixing voltage, $V_\mathrm{mix}$, as a function of magnetic field, $H$, measured in W/CoFeB microstripes for selected thicknesses, $t_\mathrm{W}$, is presented in Fig. \ref{fig:w}(a). For each $t_\mathrm{W}$, the signal is decomposed into symmetric ($V_S$) and antisymmetric ($V_\mathrm{A}$) Lorentz functions~\cite{liu_spin-torque_2011}. The dependence of the amplitudes  $V_\mathrm{S}$ and $V_\mathrm{A}$ on $t_\mathrm{W}$ is shown in Fig. \ref{fig:w}(b) and (c), together with the corresponding fitting based on the theoretical model presented in the previous section. Such an approach enables quantitative separation of the contributions from the Oersted field and interfacial spin-orbit torque to the antisymmetric part of the signal. In addition, it was found that in the case of W/CoFeB bilayer with 5-nm thick FM, the measured magnetoresistance is weakly dependent on the thickness of W, indicating AMR-like effect and negligible SMR (in contrast to thin FM case, where SMR is dominating~\cite{cho_large_2015}). As a consequence, the resulting SOT-FMR signal is not described well with the magnetoresistance change modelled with Eq.~(\ref{eq:smr}). Thus, the  experimental results were taken instead and the model was interpolated for small thicknesses of W layer. Based on Eqs.~(\ref{eq:hdl}),~(\ref{eq:hoe}), and~(\ref{eq:hso}), the evolution of the effective fields as a function of $t_\mathrm{W}$ has been determined and is presented in Fig.~\ref{fig:w}(d). In the case of W, the interface spin-orbit field is relatively weak: $\Gamma_\text{so}^{\text{W}}$ = -0.27 Oe. The experimentally determined values of $\theta^{\text{eff}}$, based on Eq.~(\ref{eq:eff_sha}), approach the fitted value of $\theta_{\text{W}}$ = -0.36 for thick W layers.
\begin{figure*}
\centering
\includegraphics[width=\textwidth]{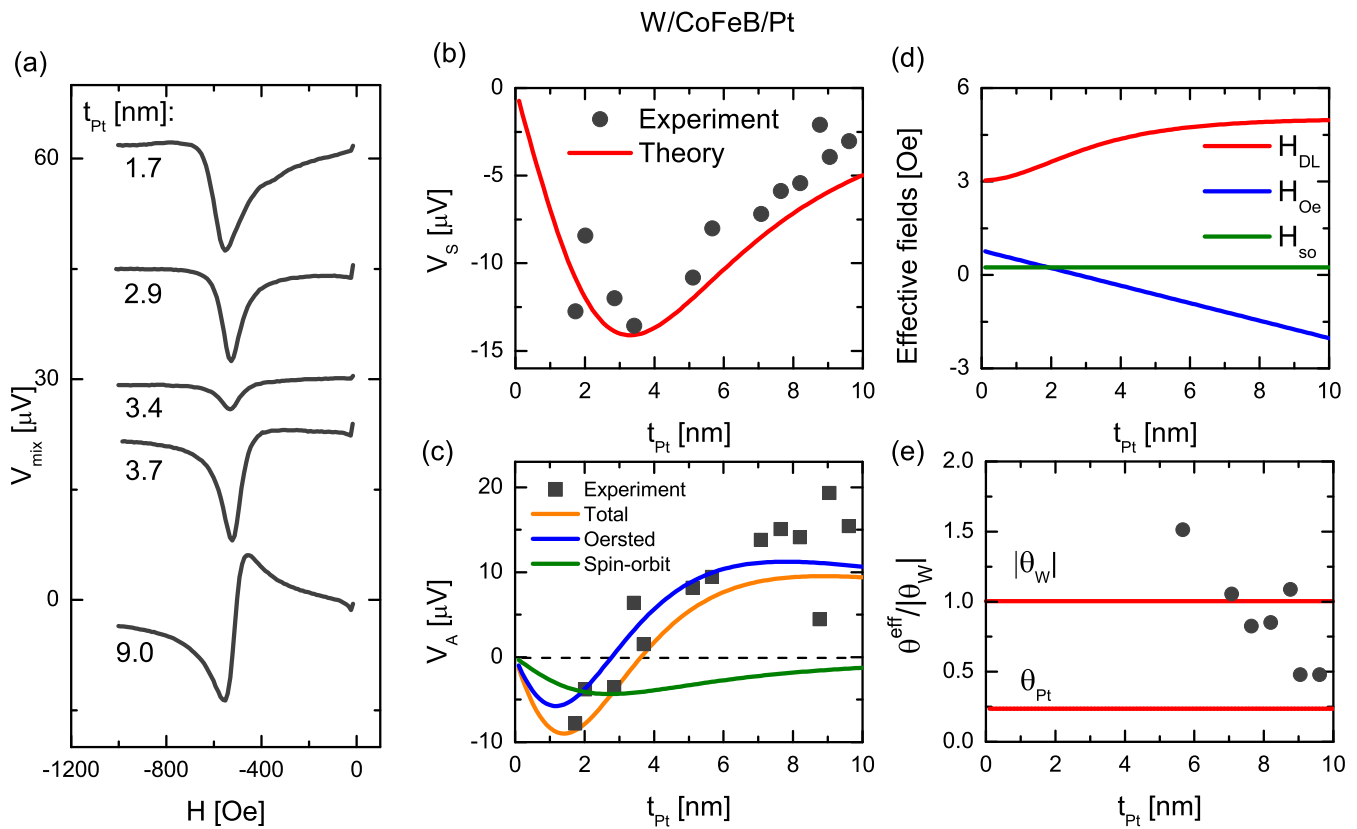}
\caption{DC mixing voltage, $V_{\text{mix}}$, measured in W/CoFeB/Pt microstripe as a function of magnetic field, $H$,  is shown in (a). The $V_{\text{mix}}$ curves for different $t_\mathrm{Pt}$ are artificially offset for clarity. Experimentally determined amplitudes: $V_\mathrm{S}$ and $V_\mathrm{A}$ are shown in (b) and (c). Solid lines represent theoretical values based on the fitting parameters obtained for bilayer systems. Components of the effective field are presented in (d). The effective spin Hall angle of the trilayer, $\theta^{\text{eff}}$, relative to $|\theta_\mathrm{W}|$, is depicted in (e). Note, that the sign of the effective spin Hall angle of the trilayer is positive.}
\label{fig:w-pt}
\end{figure*}
Similar experimental procedure as well as quantitative analysis were repeated for the CoFeB(5)/Pt($t_\mathrm{Pt}$) stripes. The corresponding results are presented in Fig.~\ref{fig:pt}. Unlike the W case, the DC mixing voltage of Pt stripes unequivocally changes sign with increasing $t_\mathrm{Pt}$, as shown in Fig. \ref{fig:pt}(a). Thickness dependence of the corresponding symmetric and antisymmetric components are shown in Fig.~\ref{fig:pt}(b) and (c), respectively. Behavior of the symmetric component is well explained by a combination of the spin Hall induced damping-like field and SMR effect, in contrast to the above described W/CoFeB bilayer. Fitting of the model to the experimental data allowed to obtain the spin Hall angle $\theta_{\text{Pt}}=0.09$, opposite in sign to the spin Hall angle obtained for W/CoFeB, and the spin diffusion length $\lambda_{\text{Pt}}=2.2~\text{nm}$. Both values agree with the data presented in the relevant literature~\cite{Rojas_Sanchez_spin_2014,zhang_role_2015,sagasta_tuning_2016}.

The sign change of the antisymmetric part of the signal occurs due to a stronger, compared to W/CoFeB bilayer, interfacial spin-orbit field, $\Gamma_\text{so}^{\text{Pt}}$ = 0.54 Oe, which dominates $V_\mathrm{A}$ for $t_\mathrm{Pt} < 3~\textrm{nm}$. Similar field-like torque contribution in Pt/Co/MgO multilayers was measured for the same Pt thickness using harmonic Hall voltage measurements~\cite{nguyen_spin_2016}.

Finally, the spin-orbit-torque-induced dynamics in W/CoFeB/Pt trilayers was investigated and the corresponding results are shown in Fig.~\ref{fig:w-pt}. Similar to the CoFeB/Pt bilayer, symmetry of the SOT-FMR signal changes for $t_\mathrm{Pt} = 3~\textrm{nm}$, as shown in Fig.~\ref{fig:w-pt}(c). However, in the case of a trilayer, the spin currents from both HM layers are absorbed in FM, which results in  an increase in $V_\mathrm{S}$. 
Note, that for $t_\mathrm{Pt}$ = 2.5 nm, the Oersted field and interfacial spin-orbit contributions from both W and Pt are minimized and therefore a pure spin current induced dynamics is observed.
Solid lines in Fig. \ref{fig:w-pt}(b) and (c) are drawn based on the fitting parameters obtained for W/CoFeB and CoFeB/Pt bilayers showing good agreement between the experimental values and theoretical predictions.

The effective spin Hall angle of the trilayer system, shown in Fig.~\ref{fig:w-pt}(e), is 1.5 times larger than that for W/CoFeB bilayer alone for $t_{\text{Pt}}\approx 6~\text{nm}$, while for thicker platinum it decreases to a value closer to the one obtained for the CoFeB/Pt bilayer. This drop can be explained by larger spin-mixing conductance at the CoFeB/Pt interface and thus larger spin current flowing through this interface.

\section{Summary}
\label{sec:summary}
In summary, the spin-orbit-torque-induced dynamics in W/CoFeB and CoFeB/Pt bilayers and W/CoFeB/Pt trilayer was investigated experimentally by the spin-orbit-torque ferromagnetic resonance technique. Both symmetric and antisymmetric parts were resolved in the SOT-FMR signals from the microstripes  investigated. Variation of the magnitudes of the corresponding signals with increasing heavy metal thickness  was fitted to the developed theoretical model. From the application point of view, it is important to note that when combining ferromagnets with materials, which exhibit strong spin-orbit coupling, such as W and Pt, interfaces between those materials play very important role in determination of the torques exerted on magnetization and other properties of the constituent materials. We have determined the magnitude of interfacial spin-orbit fields from W/CoFeB and CoFeB/Pt interfaces and shown how they influence the spin Hall angle and spin diffusion length in these bilayers as well as in W/CoFeB/Pt trilayer. In particular, we have shown that for specific thicknesses of the Pt and W layers, the Oersted field is cancelled by the interfacial spin-orbit field, which leads to a pure spin-current induced dynamics.

\section*{Acknowledgments}
The authors would like to thank J. Ch\k{e}ci\'{n}ski for help in calculations and M. Schmidt and J. Aleksiejew for technical support. This work was supported by the National Science Centre, Poland, grant No. UMO-2015/17/D/ST3/00500. \L{}.~K., S.~\L{}., K.~G., and T.~S. acknowledge support from National Science Centre research project 2016/23/B/ST3/01430 (SPINORBITRONICS). Microfabrication was performed at Academic Center for Materials and Nanotechnology of AGH University.

\end{document}